# Heavy Hole vs. Light Hole Spin Qubits: A Strain-Driven Study of SiGe/Ge and GeSn/Ge

Kelvin Dsouza,[1] Patrick Del Vecchio,[2] Nicolas Rotaru,[2] Oussama Moutanabbir,[2] and Daryoosh Vashaee[1,3a]

AFFILIATIONS
[1]Electrical and Computer Engineering Department, North Carolina State University, Raleigh, North Carolina 27616, USA
[2]Department of Engineering Physics, École Polytechnique de Montréal, Montréal, Québec H3C 3A7, Canada
[3]Materials Science and Engineering Department, North Carolina State University, Raleigh, North Carolina
a) Author to whom correspondence should be addressed: dvashae@ncsu.edu

**ABSTRACT**

This work investigates and compares the impact of strain on heavy hole (HH) spin qubits in SiGe/Ge and light hole (LH) spin qubits in GeSn/Ge heterostructures, focusing on energy states, g-factor, Rabi frequency, spin relaxation, and dephasing times. By exploring the distinct properties of HH and LH spin qubits under strain, we demonstrate how strain serves as a tunable parameter to optimize qubit performance. The study highlights that LH spin qubits in Ge quantum dots exhibit lower relaxation rates and higher Rabi frequencies, offering significant advantages for addressing current challenges in gate-defined spin qubits. A significant difference is observed in the g-factor anisotropy, where for HHs the out-of-plane g-factor ($g_\perp$) is larger than the in-plane g-factor ($g_\parallel$), whereas for LHs, the in-plane g-factor dominates both in GeSn/Ge and SiGe/Ge quantum dots. This comparative analysis provides a deeper understanding of HH and LH spin dynamics, advancing the development of scalable quantum technologies based on strained Ge systems.

## I. INTRODUCTION

Quantum computing has emerged as a transformative field, offering the potential to solve problems intractable for classical systems. Among the various quantum bit (qubit) platforms, spin qubits have gained significant attention due to their compactness, scalability, and compatibility with existing semiconductor technologies. Electron spin qubits, formed by confining single electrons in quantum dots (QDs), particularly in Si and GaAs, have demonstrated long relaxation times and high coherence essential for robust quantum operations[1,2]. However, their performance is fundamentally hindered by challenges such as decoherence induced by hyperfine interactions with nuclear spins, spin-phonon coupling, and valley degeneracy in Si-based gate-defined quantum dots. Valley splitting, in particular, depends on atomistic device features and is challenging to control across large-scale quantum dot arrays, limiting the scalability of electron spin qubits[3-7].

Recent studies have explored strategies to address these challenges through innovative approaches in QD systems. For example, dynamic susceptibility and quantum Fisher information analysis have revealed the intricate interplay between exchange interactions and entanglement in quantum dots, offering insights into optimizing their coherence and functionality for quantum technologies.[8] Additionally, novel cooling mechanisms based on microwave-induced depopulation and phonon filtering have been proposed to minimize thermal noise around QD qubits, effectively reducing their local temperatures to below 10 mK at bath temperatures of 1 K.[9] These advancements lay the groundwork for developing more robust and scalable qubit architectures.

To overcome the specific limitations of electron spin qubits, hole spin qubits have emerged as a promising alternative. These qubits leverage the distinct properties of holes in semiconductors, addressing the shortcomings of electron spin qubits while offering unique advantages. Hole spin qubits are free from the valley degeneracy challenges that hinder the scalability of Si-based electron qubits. Additionally, the strong spin-orbit coupling (SOC) in holes enables electric field-based spin manipulation, simplifying device architecture by eliminating the need for micromagnets. The p-orbital character of holes suppresses hyperfine interactions, reducing decoherence and enhancing qubit coherence times. Due to the p-character of their orbital wave functions, hole spins experience negligible contact hyperfine interaction, leaving only the relatively weak orbital and dipolar hyperfine couplings. Isotopic enrichment can further reduce, and potentially eliminate, the number of nuclear spins interacting with a hole spin qubit.

Consequently, the primary decoherence channels for hole spins are governed by spin-orbit interaction (SOI), which exposes the spin qubit to fluctuations in the electrical environment. For conduction electron spins, the SOI depends linearly on kinetic momentum and, at the lowest order, primarily leads to spin relaxation, as the resulting effective magnetic field is always perpendicular to the spin qubit's quantization axis. In contrast, hole spins exhibit a more complex SOI, influenced by factors such as heavy-hole-light-hole mixing and strain distribution in the material. This interaction can have both linear and cubic dependencies on kinetic momentum, among other contributions, adding complexity to the dynamics of hole spin qubits.

Strained germanium (Ge) quantum wells have demonstrated significant promise for implementing hole spin qubits, with properties such as high spin-0 isotope abundance, reduced effective mass, and the absence of valley degeneracy. Strain engineering in Ge-based heterostructures enables precise control over heavy hole (HH) and LH states, with the latter offering distinct advantages for quantum computing applications. These include high Rabi frequencies for rapid spin manipulation and efficient spin-photon interfaces for quantum communication. Compared to Si-based systems, Ge also allows for larger quantum dots, relaxing fabrication constraints and improving device scalability[10-14].

The band structure of Ge under tensile and compressive strain is a crucial factor in its application for quantum computing. Density Functional Theory (DFT) calculations reveal that Ge exhibits a degeneracy at 0% strain in the bulk, which disappears with increasing or decreasing strain. Under compressive strain, HH states dominate, while LH states dominate under tensile strain. In quantum dots, confinement leads to quantized energies, with the crossing point of HH and LH states occurring at specific strain levels. The crystal becomes highly anisotropic under strain, showing degeneracy in the xy plane but not in the z plane. These properties enable the tuning of qubit states and their interactions with external fields, essential for scalable quantum processor designs. The schematic band structure of strained Ge under compressive and tensile strain is illustrated in Figure 1, which is the consequence of the transitions from cubic to tetragonal symmetry along the z-axis.[11] This figure highlights the degeneracy at 0% strain in the bulk and its disappearance with strain variations. Compressive strain raises the HH states in the valence band, while tensile strain elevates the LH states, enabling LH-dominated properties in tensile-strained Ge.

An important metric in evaluating spin-based qubit systems is the coherence time of the spin in quantum dots. Long coherence times are essential for implementing quantum algorithms and error correction schemes. The decay time, characterized by the spin relaxation time ($T_1$), is usually associated with the spin-flip time. The spin decoherence time ($T_2$) represents the lifetime of a superposition of spin-up and spin-down states. Another critical factor is spin dephasing time ($T_2^*$), which is influenced by charge noise arising from local electric field fluctuations. While strong SOC in holes provides advantages for spin manipulation, it also introduces a pathway for charge noise to affect qubit coherence. This can be mitigated through g-factor sweet spots, accessible via strain or magnetic field orientation, rendering qubit energy levels less sensitive to electric field fluctuations. Advances in material quality, such as isotopically purified Ge, and refined heterostructure designs further contribute to enhancing qubit coherence and reducing environmental noise.

Emerging GeSn/Ge heterostructures provide a novel platform for hole spin qubits, enabling LH ground state confinement at lower strain levels compared to SiGe/Ge systems. The GeSn/Ge configuration is particularly advantageous for light hole confinement, which facilitates coherent spin control without an external magnetic field and provides efficient spin-photon interfaces. However, a detailed understanding of their spin dynamics, coherence properties, and response to strain remains an open research challenge. Such insights are critical for optimizing key performance metrics such as the g-factor, Rabi frequency, decoherence, and dephasing times, which directly influence qubit fidelity and scalability.

In recent years, hole spin qubits have made significant progress in single spin manipulation, spin readout, scalability, and interfacing with existing superconducting circuitry, showing potential to surpass electron spin qubits. This paper provides an in-depth analysis through ab initio modeling of strained Ge systems in SiGe/Ge and GeSn/Ge configurations, focusing on heavy hole and light hole states. We utilize k·p. theory to exploit crystalline symmetries, simplifying the calculations of the quantum dot wave functions. Our study evaluates the potential of strained Ge-based systems to address the limitations of electron spin qubits and advance the development of scalable quantum processors. Additionally, we investigate the system behavior based on operations such as electric dipole spin resonance (EDSR) and calculate the Rabi frequencies based on the change in strain on the device. We also evaluate the decoherence and dephasing of the qubit system, taking into account spin-orbit coupling and charge noise, respectively. Our primary objective is to compare the performance of SiGe/Ge and GeSn/Ge heterostructures, emphasizing their potential for scalable hole spin qubit implementations. Through the integration of first-principles modeling and practical heterostructure designs, this work seeks to determine the optimal material and strain configurations to advance hole spin QD qubit technology.

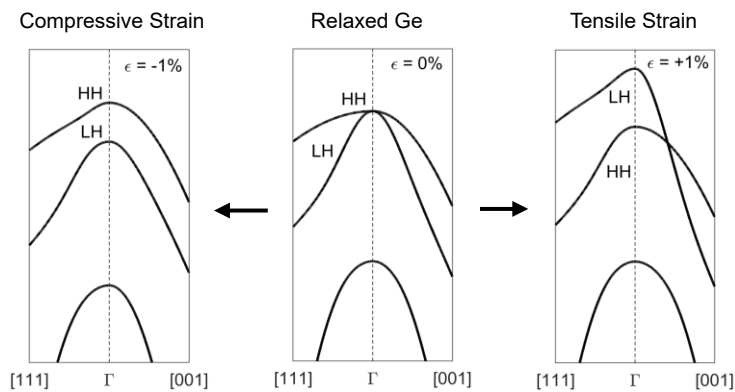

Figure 1: Schematic band structure of strained Ge under compressive and tensile strains. The transitions from cubic to tetragonal symmetry along the z-axis modify the electronic structure, lifting degeneracies at 0% strain in the bulk. Compressive strain elevates HH states, while tensile strain favors light hole LH states in the valence band, enabling strain-tunable qubit properties.

## II. METHODOLOGY

To compare the performance of two Ge-based heterostructure technologies for spin qubits, we consider planar QW configurations with the following compositions: (1) $Si_{0.2}Ge_{0.8}$/Ge/$Si_{0.2}Ge_{0.8}$ [12,13], and (2) $Ge_{0.85}Sn_{0.15}$/Ge/$Ge_{0.85}Sn_{0.15}$[14]. These heterostructures were selected to explore the distinct properties of hole spin qubits formed in strained Ge systems and to evaluate the impact of strain engineering on key qubit performance metrics.

In the SiGe/Ge/SiGe heterostructure, a pure Ge layer (16 nm) is epitaxially grown between relaxed $Si_{0.2}Ge_{0.8}$ layers (50 nm). In contrast, the GeSn/Ge/GeSn configuration consists of a 16 nm pure Ge layer sandwiched between $Ge_{0.85}Sn_{0.15}$ layers (50 nm) strained to 1.5% compressive strain. In both systems, the central Ge layer serves as the quantum well, where the two-dimensional hole gas (2DHG) is formed. Lateral confinement, induced by the potential applied across the metal electrodes, traps the holes in a quasi-one-dimensional space, resulting in the formation of hole quantum dots.

The metal electrodes are deposited on an $Al_2O_3$ capping layer, which serves as an insulating barrier, enabling precise control over the confinement potential. The configuration allows for the application of both DC and microwave signals, facilitating quantum dot confinement and EDSR-based spin manipulation. A schematic diagram illustrating the two heterostructure designs is presented in Figure 2.

In the SiGe/Ge/SiGe structure, the relaxed SiGe layers induce compressive strain in the central Ge layer due to the larger lattice constant of Ge compared to Si. This compressive strain elevates the HH states in the Ge QW, which are already dominant due to quantum confinement effects. To comprehensively analyze the effects of strain, we also consider hypothetical tensile strain scenarios in the SiGe/Ge/SiGe system. This allows us to explore both HH and LH states under varying strain conditions, providing a broader understanding of their impact on qubit properties.

In the GeSn/Ge/GeSn configuration, relaxed layers do not induce hole confinement in the Ge layer. However, the Ge layer can experience varying degrees of tensile strain depending on the Sn content in the GeSn layers, which have a larger lattice constant. This tensile strain raises the LH states, and beyond a critical strain value, the valence band alignment facilitates the formation of an LH QW. At this point, an LH spin qubit can be realized through gate-induced confinement.[15-17]

The study employs both ab initio modeling using Quantum Espresso[18] and semi-empirical k·p theory using the QTCAD software[19] to analyze the quantum dot behavior in these heterostructures. DFT is used to evaluate the electronic band structure and strain-dependent properties of the materials, while k·p theory simplifies the calculations of QD wave functions by exploiting crystalline symmetries. Key parameters such as effective mass, deformation potentials, and Luttinger parameters are derived from DFT and incorporated into the k·p Hamiltonian to calculate the eigenstates of the system. The effects of strain, gate bias, and electrode configurations on the QD's confinement potential and energy levels are systematically examined.

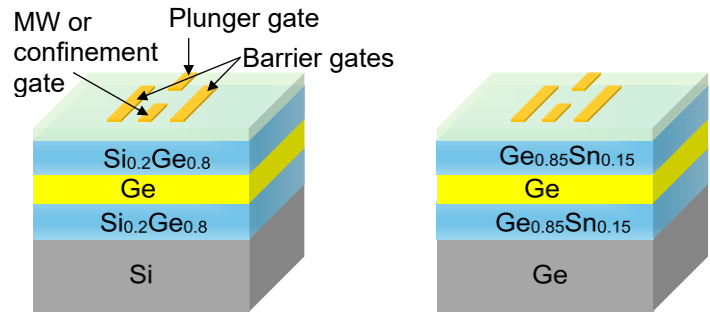

Figure 2: Schematic of gate-defined SiGe/Ge/SiGe and GeSn/Ge/GeSn QD qubit devices. The four gate electrodes are labeled to illustrate their roles: the plunger gate (top) controls the QD's potential and occupancy; the left and right barrier gates regulate tunneling between the QD and adjacent reservoirs; and the confinement or microwave gate (bottom) defines the QD's spatial extent and facilitates microwave-induced EDSR for spin manipulation. In the SiGe/Ge system, a relaxed SiGe layer is epitaxially grown on a Si substrate, while in the GeSn/Ge structure, a graded strained GeSn layer is grown on Ge substrates. The central Ge layer is strain-engineered to achieve various strain levels, enabling precise control over the energy states confined within the QD.

To incorporate strain into the k·p Hamiltonian, the Luttinger-Kohn-Bir-Pikus (LKBP) Hamiltonian is solved[20,21]. For this purpose, we utilize the QTCAD software suite[22-24], which employs a 4-band model to calculate the electronic properties under various system parameters. This approach enables accurate characterization of the electronic behavior of holes in QDs by integrating first-principles and semi-empirical methods. The k·p calculations allow differentiation between the contributions of HH and LH to the quantum states.

To achieve QD confinement, a depletion-mode design is employed, where gate electrode patterns are deposited on the top layer. A uniform bias is applied to all gate and plunger electrodes, depleting the hole density in the confinement region[25]. The applied potential induces band bending due to the electric field across the device, and precise control over the potential is essential for confining holes within the QW.

The first step involves solving the Poisson equation across the device to determine the electrostatic potential and band alignment within the heterostructure. This calculation initializes the electronic environment of the system, ensuring accurate input parameters for the subsequent Hamiltonian solution. Next, the Schrödinger equation is solved within the quantum dot subregion using k·p theory. The results of these simulations are illustrated in Figure 3, which shows the electrostatic potential and the resulting ground-state wavefunction.

## III. RESULTS AND DISCUSSIONS

### III.I. Confinement Study

The spatial distribution and character of localized wavefunctions in QDs provide crucial insights into carrier confinement and behavior under various strain conditions. Figure 3(a) illustrates the electrostatic potential under an applied gate bias for both the SiGe/Ge and GeSn/Ge heterostructures. In this analysis, the Ge layer in the SiGe/Ge structure is assumed to be relaxed, while the Ge layer in the GeSn/Ge heterostructure is under 2.2% tensile strain. The results demonstrate optimal confinement within the gate electrodes in both SiGe/Ge and GeSn/Ge heterostructures, as shown in Figure 3 (b). Figure 3 (c) and (d) show the confining potential in the plane of QW for the two heterostructures.

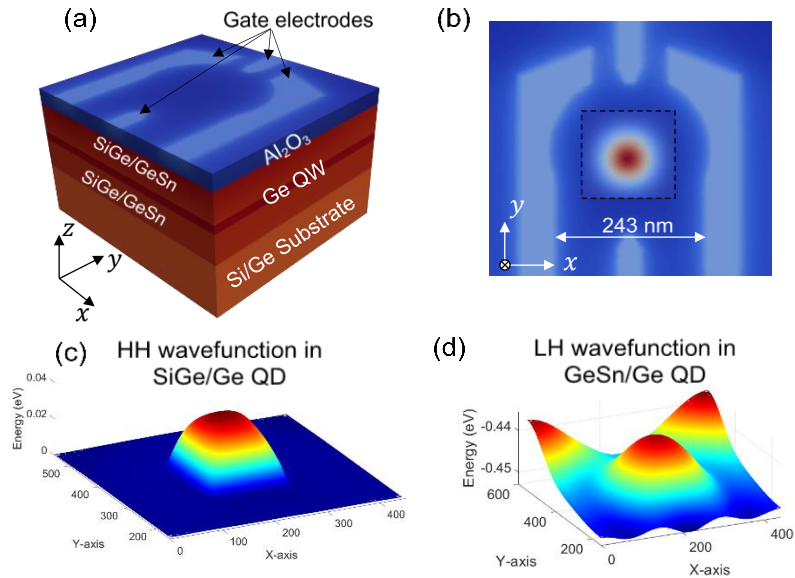

Figure 3: The electrostatic potential is obtained by solving the Poisson equation across the full 3D device structure. (a) shows the potential at the metal electrode, where a bias is applied to define the QD region. (c) overlays the QD's wave function, demonstrating its spatial location with respect to the location of the gate electrodes. The Hamiltonian is solved for the subdomain identified with a dashed square. (c) and (d) illustrate the potential energy within the Ge QW layer for SiGe/Ge with relaxed Ge layer and GeSn/Ge with Ge at 2.2% tensile strain configurations.

Figure 4(a) illustrates the strain-dependent confinement along the z-direction within the QW HH and LH states in a SiGe/Ge/SiGe QD heterostructure. At a compressive strain of -1.0%, HHs occupy the ground state, while LHs are transferred to excited states. This configuration persists until a critical strain threshold is reached. Beyond this threshold, a transition occurs where LHs descend to occupy the ground state. This strain-induced transition demonstrates the tunability of the hole ground state character in Ge QDs.

The 3D visualizations in Figure 4(a) elucidate the spatial confinement characteristics of both HH and LH wavefunctions within the QD as a function of strain. Interestingly, both LH and HH wavefunctions show strong confinement, remaining localized within the QD boundaries without significant penetration into the potential barriers. The extent of wavefunction penetration into the barriers influences interaction with defects and nuclear spins in the barrier, which could impact the coherence time, carrier lifetimes, and inter-dot coupling in multi-QD systems. Furthermore, the strain-dependent behavior of hole wavefunctions has significant implications for the electronic and optical properties of Ge-based QDs. The transition from HH- to LH-dominated ground states with increasing tensile strain can lead to modifications in the selection rules for optical transitions, potentially affecting the polarization and intensity of emitted light.

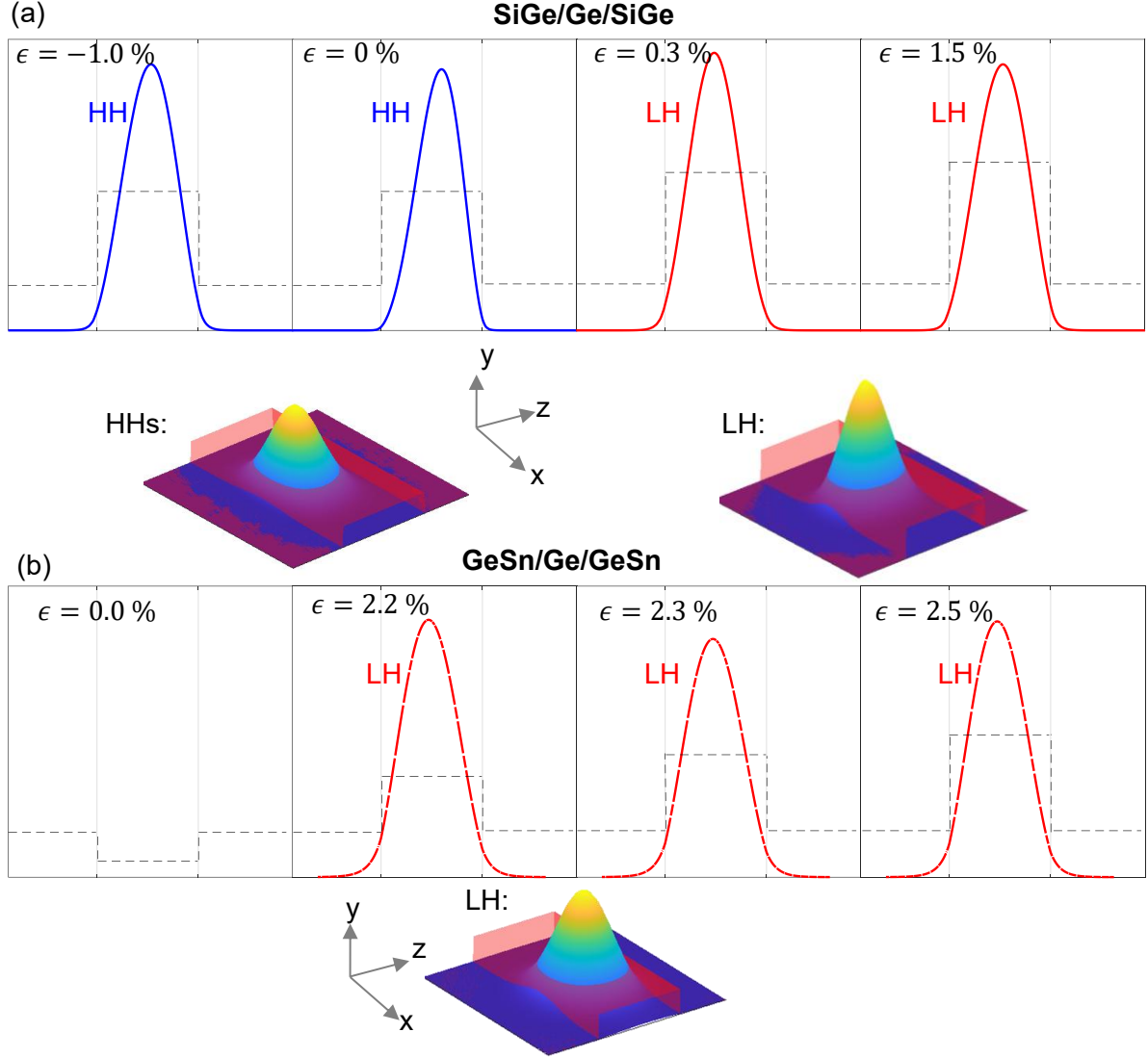

Figure 4: HH and LH wavefunctions in QDs under varying strain conditions, calculated using k·p theory at zero magnetic field (B = 0). 1D plots show the amplitude of the wavefunction in the z direction across the center of the QD, and the 3D representation shows the wavefunction in the middle of the Ge layer. Panel (a) illustrates the SiGe/Ge/SiGe heterostructure with changing strain in the Ge layer. Under compressive strain, HHs (blue) occupy the ground state energy. As strain in the Ge layer increases beyond 0.3%, LHs (red) transition to the ground state, indicating that increased tensile strain favors LH confinement in the ground state. Panel (b) depicts the GeSn/Ge/GeSn heterostructure. At compressive or lower tensile strain, holes are not confined. When tensile strain exceeds 2.2%, LHs (red) become confined within the QD and occupy the ground state, suggesting no HH confinement within the QD domain under tensile strain.

In the case of GeSn/Ge/GeSn heterostructures with a 16 nm Ge layer (Figure 4 (b)), wavefunction localization is strongly dependent on the applied strain. The energy levels remain constant until a critical tensile strain of approximately 2.2% is reached, at which point confinement occurs. This behavior arises from strain-induced modifications in the band alignment between the GeSn and Ge layers. At 0% strain, no wavefunction localization is observed within the QD. However, as tensile strain increases beyond the critical threshold, LH become confined within the QD and occupy the ground state, demonstrating the tunability of hole states in Ge-based QD structures.

A key distinction between GeSn/Ge and SiGe/Ge heterostructures lies in the spatial distribution of LH and HH wavefunctions under high tensile strain. In GeSn/Ge structures, LH wavefunctions are localized within the QD, while HH wavefunctions move to the boundaries and extend beyond the dot region. This spatial separation reduces HH-LH mixing, as overlap between their wavefunctions occurs predominantly outside the QD. In contrast, in SiGe/Ge structures, both LH and HH wavefunctions can be confined within the QD depending on the strain conditions. This structural difference has significant implications for the coherence properties of the system, particularly when the LH and HH energy levels are in close proximity. The increased spatial overlap of LH and HH states within the QD in SiGe/Ge structures can lead to enhanced mixing and coupling between these states. Consequently, this can result in a higher density of available states for scattering processes, potentially impacting the coherence time of the hole spin. The interplay between strain, spatial confinement, and wavefunction overlap thus plays a crucial role in determining the quantum coherence properties of these heterostructures.

The GeSn/Ge structure predominantly exhibits LH character under tensile strain. However, in the SiGe/Ge system, achieving LH character necessitates the application of tensile strain levels that exceed what is currently experimentally feasible for this particular structure. The critical strain for transitioning from HH to LH ground states depends on the Ge layer thickness and approaches the bulk value of 0.0% as the thickness increases. For a SiGe/Ge/SiGe structure with a 16 nm Ge thickness, band mixing occurs at a critical strain of approximately 0.28%. The energy states for LH and HH as a function of in-plane strain are shown in Figure 5.

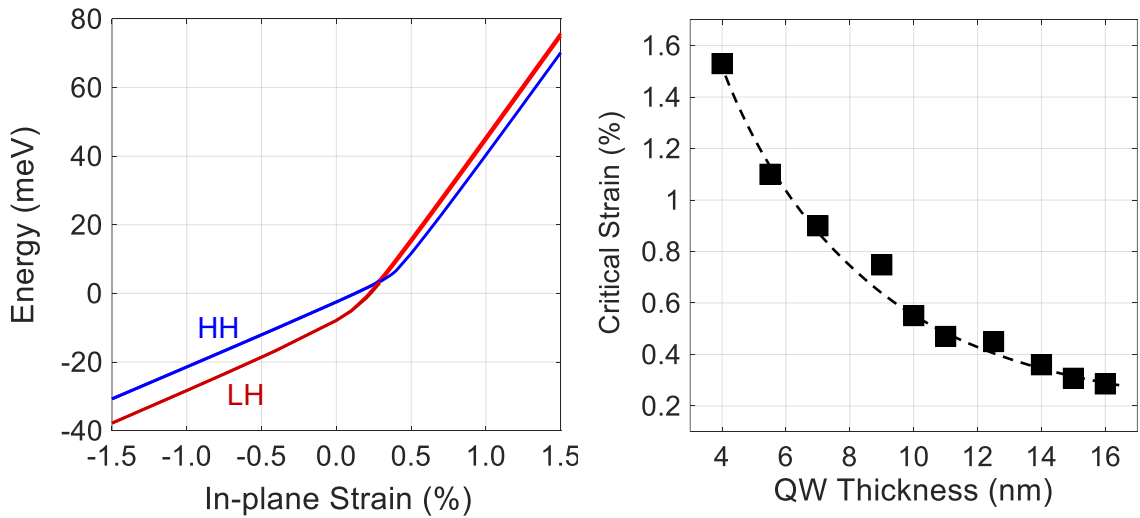

Figure 5: (a) Energy levels as a function of strain for LH and H states in a 16 nm thick SiGe/Ge/SiGe QD. The critical strain, where LH and HH states interchange their energy levels, is clearly identified. (b) Critical strain as a function of Ge thickness, illustrating an exponential decrease in critical strain with increasing thickness. At larger thicknesses, the critical strain approaches the bulk value of 0%, where LH and HH states become degenerate. is too small to effectively couple to electric fields.

The relationship between critical strain and Ge thickness reveals that thinner layers require higher tensile strain to achieve LH ground states. As the Ge layer thickness increases, the system begins to resemble a bulk-like structure with degenerate LH and HH states at zero strain. This behavior is illustrated in Figure 5(b), where confinement strain is plotted against Ge thickness, showing that thicker layers approach bulk behavior.

### III.II. *g*-FACTOR

The top of the valence band of relaxed Ge at the Γ point is fourfold degenerate. This degeneracy is lifted by spin-orbit coupling and further split by the broken symmetry introduced by applied strain. Additionally, the application of a magnetic field induces Zeeman splitting in the degenerate LH and HH bands, with the magnitude and nature of the splitting depending on the direction of the magnetic field and the confinement potential.

The *g*-factor, a critical parameter for qubit performance, is typically calculated using a perturbative approach such as the Schrieffer-Wolff transformation.[26] In this study, we calculate the g-factor by applying magnetic fields parallel and perpendicular to the plane of the QW, determining the energy splitting between states and using the resulting energy differences to calculate the g-factor directly. The out-of-plane g-factor ($g_\perp$) characterizes the spin's response to a magnetic field normal to the QW plane, while the in-plane g-factor ($g_\parallel$) describes the spin's response to a magnetic field within the QW plane.

In the SiGe/Ge heterostructure, the *g*-factor is evaluated for both $g_\parallel$ and $g_\perp$ for HH and LH states confined in the QD, as shown in Figure 6. As discussed in the Appendix, $g_\parallel$ is isotropic within the xy-plane. For magnetic fields applied in the z-direction, the field predominantly couples to HH states, resulting in a high $g_\perp$ under compressive strain. At the critical strain, the LH states become the ground state, significantly lowering the *g*-factor for HH states. As depicted in Figure 6, HH states exhibit negligible coupling to in-plane magnetic fields, resulting in a $g_\parallel$ that remains consistently low, around 0.2, across all compressive strain values.

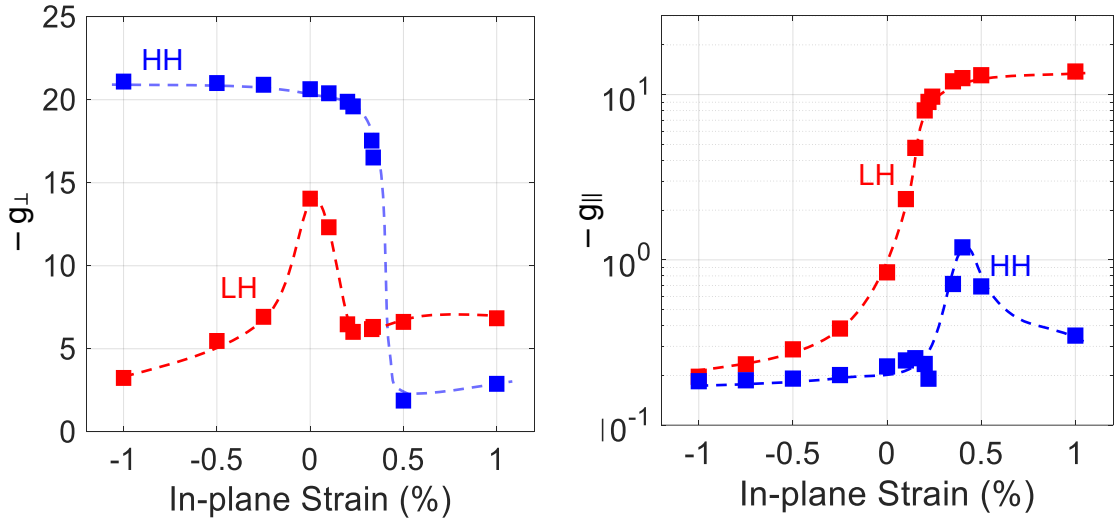

Figure 6: The in-plane ($g_\parallel$) and out-of-plane ($g_\perp$) g-factors calculated for the Ge QD in a SiGe heterostructure. The $g_\perp$ for HH (blue) remains constant under compressive strain up to the critical strain, after which it drops significantly as LH occupy the ground state. The $g_\parallel$ for HH remains near 0.2 throughout but exhibits a small peak at the critical strain due to HH-LH mixing. For LH (red), the $g_\perp$ increases under compressive strain, peaking near the critical strain due to band mixing, and stabilizes at a g-factor of 7. The $g_\parallel$ for LH exhibits a high value under tensile strain beyond the critical strain, where LH dominates as the ground state.

HH states show negligible coupling to in-plane magnetic fields resulting in a small $g_\parallel$ that remains close to 0.2 across all negative strain values. This is mainly because their interaction is limited to the linear

$J$-operator in the effective Hamiltonian.[27] Interestingly, a small peak appears for $g_\parallel$ near the critical strain. This is attributed to cubic $g$-factor contributions arising from the mixing of HH and LH states.

In contrast, the in-plane magnetic field couples strongly to the LH, with $g_\parallel$ increasing with strain, eventually saturating at a value of 14. The peak observed near the critical strain can be attributed to the mixing of HH and LH wavefunctions. The proximity between LH and HH states can create sweet spots where qubit sensitivity to noise is minimized due to enhanced tunability of the g-factor or reduced sensitivity to electric field fluctuations.[27,28]

In the GeSn/Ge system, where HH states are not confined, we focus on the $g$-factor for LH under various strain conditions. The $g_\perp$ for LH is a weak function of strain, stabilizing at a value close to 7. Meanwhile, $g_\parallel$ grows consistently with increasing strain, reaching a value of 13.3 at a strain level of 2.5%.

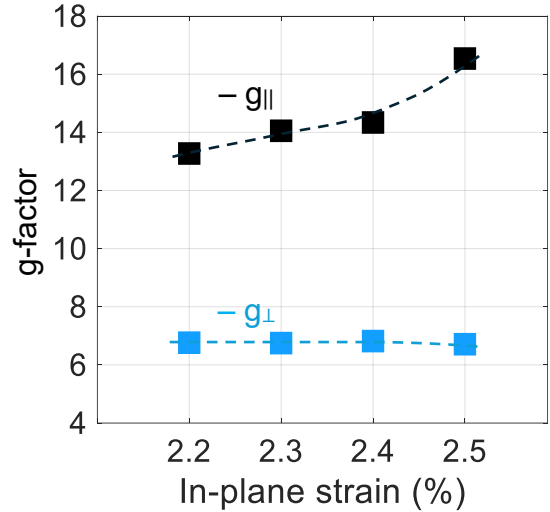

Figure 7: The LH $g$-factor in a GeSn/Ge QD. The in-plane g-factor ($g_\parallel$) increases with the applied strain as it couples with the magnetic field, reaching a value of 16.5 at 2.5% biaxial strain. In contrast, the out-of-plane g-factor ($g_\perp$) remains relatively constant at approximately 6.8 despite the increasing strain.

### III.III. Comparison of g-Factor Anisotropy in LH and HH Spin Qubits

In SiGe/Ge systems, HHs exhibit a high $g_\perp \approx 21$, which directly determines the Zeeman energy splitting ($E_z = g_\perp \mu_B B_z$). On the other hand, LHs in GeSn/Ge systems show a lower $g_\perp \approx 7$, which reduces the Zeeman splitting under similar field conditions. $g_\perp$ in both HH and LH spin qubits shows minimal variation with strain, as observed in Figure 6 and Figure 7. This suggests that the tunability of $g_\perp$ with strain is limited in the studied range, which may reduce the ability to fine-tune the qubit properties via strain engineering.

For EDSR, HHs in SiGe/Ge exhibit a very small $g_\parallel \approx 0.2$, indicating reduced sensitivity to SOC effects, negligible coupling to in-plane electric fields, and limited suitability for electric-field-based manipulation. The minimal $g_\parallel$ helps suppress spin-phonon coupling in the presence of strain and field-induced anisotropy, reducing decoherence rates. However, the inability to efficiently manipulate HH spin states using electric fields necessitates the use of micromagnets or external magnetic field gradients for spin control, complicating device architecture. In contrast, LHs in GeSn/Ge demonstrate a significantly larger $g_\parallel > 10$, enabling strong coupling to in-plane electric fields and efficient EDSR. This capability eliminates the need for additional magnetic fields in the xy-plane and simplifies device design by reducing the reliance on micromagnets, although micromagnets may still be advantageous for generating field gradients to enhance control. The large $g_\parallel$ also increases susceptibility to spin-phonon interactions, particularly under strain, leading to higher rates of relaxation and decoherence. Optimizing $g_\parallel$ is crucial to balancing spin manipulation efficiency with coherence.

The dominance of $g_\perp$ over $g_\parallel$ (with $g_\parallel \ll g_\perp$) indicates reduced sensitivity to SOC effects, making HH spin qubits less susceptible to phonon-induced relaxation and decoherence. This stability enhances the coherence times of HH spin qubits but limits their electric-field-based spin manipulation, as $g_\parallel$ is too small to effectively couple to electric fields.

The trade-offs between HH and LH spin qubits are evident when considering coherence versus manipulation. HH spin qubits benefit from higher coherence times due to their reduced coupling to SOC and phonon interactions. However, their limited $g_\parallel$ constrains electric-field-based manipulation, requiring additional hardware complexity. In comparison, LH spin qubits offer superior electric-field tunability and faster Rabi oscillations through EDSR, owing to their high $g_\parallel$. This comes at the cost of increased phonon-induced decoherence, which requires careful engineering of strain and material properties. Additionally, HH spin qubits in SiGe/Ge are less sensitive to strain and field-induced anisotropy due to their smaller $g_\parallel$, making them more suitable for environments with minimal strain control. LH spin qubits in GeSn/Ge, on the other hand, are highly tunable via strain, allowing fine control over their g-factor components. This tunability, however, also makes them more vulnerable to decoherence caused by strain fluctuations and electric field noise.

### III.IV. EDSR STUDY

To operate the quantum dot system as a qubit, an external magnetic field is applied to establish the spin qubit's quantization axis. EDSR is implemented by treating the system as a two-level model under the rotating wave approximation. A microwave driving signal is applied through the top electrode, which generates an oscillating electric field. This time-varying field interacts with the hole's charge and spin, causing it to oscillate within the quantum well. Through SOC, this interaction drives Rabi oscillations, allowing coherent manipulation of the spin state.

SOC in quantum wells is primarily governed by two mechanisms: Rashba spin-orbit coupling (RSO), which originates from structural inversion asymmetry[29], and Dresselhaus spin-orbit coupling (DSO), which arises from bulk inversion asymmetry[30]. In group IV materials, the *k*-linear terms of SOC are absent due to their inversion symmetry. In Ge, near-perfect inversion symmetry eliminates the Dresselhaus mechanism, leaving the k-cubic Rashba SOC as the dominant contributor. This SOC leads to transitions involving $\Delta n = \pm 3$, which complicates the $\Delta n = \pm 1$ transitions typically required for efficient EDSR[31,32]. Thus, the spin-orbit coupling in Ge under an electric field is primarily attributed to the k-cubic Rashba term[10].

In this study, a gate bias of 1 V is applied to the leads of the system, along with an AC driving field of $E_{AC}$ = 9.6mV/μm at a magnetic field B=1 T. EDSR driven by SOC has been experimentally demonstrated in planar Ge quantum dots with driving frequencies exceeding 100 MHz[10,33-35]. Notably, the Rabi frequency has been shown to increase linearly with the amplitude of the applied AC field in strained Ge quantum wells[36].

A critical aspect of this study is the impact of strain on the Rabi frequency, illustrated in Figure 8. In the case of a SiGe/Ge quantum dot, the Rabi frequency increases with applied strain, likely due to enhanced band mixing near the critical strain, where HH and LH states interact strongly. This interaction amplifies the SOC, facilitating transitions between spin states under the driving field. At the critical strain, the Rabi frequency peaks at approximately 500 MHz, reflecting the maximum contribution from HH-LH mixing. Beyond the critical strain, as the system transitions into the LH-dominated regime, the Rabi frequency stabilizes at 100 MHz, where the effects of SOC and *g*-factor anisotropy reach equilibrium.

Conversely, in a GeSn/Ge quantum dot, LH states dominate due to the induced tensile strain in the system, which reduces HH-LH mixing. As a result, the Rabi frequency changes slowly with strain and remains around 0.2 to 1 GHz across varying strain levels. This behavior highlights the reduced sensitivity of GeSn/Ge quantum dots to strain-induced band mixing. For LH qubits, higher Rabi frequencies are often associated with increased SOC strength, which can result in shorter relaxation and coherence times.

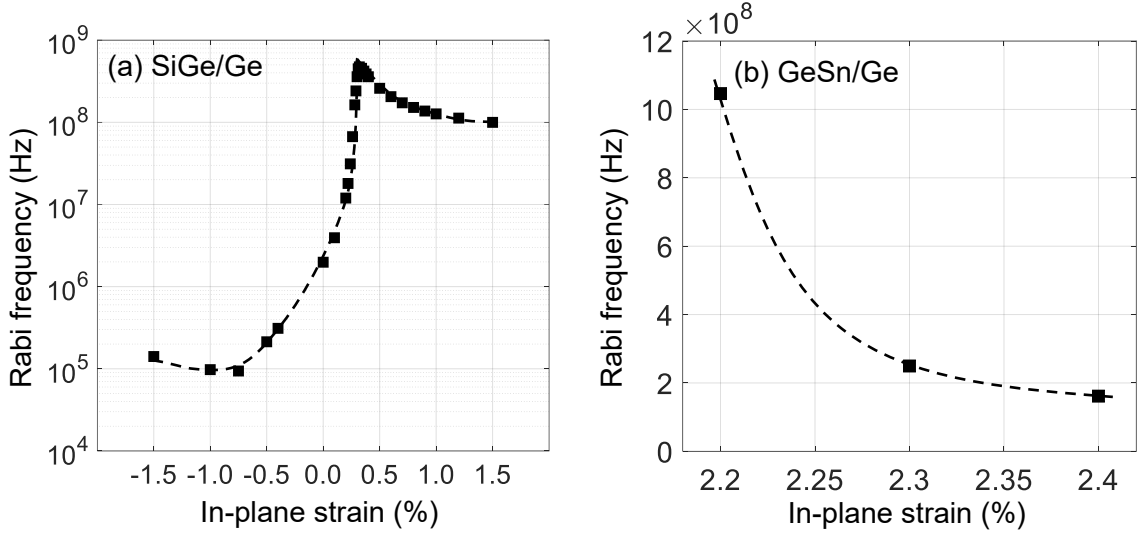

Figure 8: Rabi frequencies vs. strain for $E_{AC}$=9.6 mV/μm at B=1 T. (a) SiGe/Ge quantum dot: The Rabi frequency increases with strain, peaking near the critical strain due to HH-LH mixing and stabilizing as it transitions into the LH-dominated regime. (b) GeSn/Ge quantum dot: The Rabi frequency reduces with strain remaining in the range of 0.2 to 1 GHz.

### III.V. SPIN RELAXATION TIME

While SOC is a key enabler for spin manipulation in strained Ge, it also contributes to high spin relaxation rates. Spin relaxation in holes occurs primarily through SOC-mediated coupling to bosonic phonons, specifically the acoustic phonon modes of Ge. The SOC affects the wavefunctions more significantly than the energy levels, as the latter are only influenced in the second order. This implies that SOC induces mixing of the wavefunctions, which facilitates spin relaxation.

Decoherence in the QD system is calculated by analyzing phonon-induced relaxation due to the interaction between holes and acoustic phonons. Optical phonons do not contribute significantly to relaxation in this low-energy regime. Given the absence of Dresselhaus and k-linear Rashba terms in group IV materials, only the cubic Rashba term is considered in this study. Spin relaxation induced by HH through phonon-HH interactions follows the formalism developed by Bulaev et al.[37,38]. The phonon-induced spin relaxation rate is given by:

$$W_{1n}^R = \frac{\alpha^2 \omega_z^7}{2^8 \hbar^3 \pi^2 \rho \Omega^6} \left(N_{\omega_z} + 1\right) \left(\frac{\omega_+^3}{3\omega_+^3 + \omega_z} - \frac{\omega_-^3}{3\omega_-^3 - \omega_z}\right)^2 \sum_\alpha s_\alpha^{-9} e^{-\omega_z^2 l^2 / 2s_\alpha^2} I^{(7)} \quad (1)$$

Where,

$$I^{(k)} = \int_0^{2\pi} d\varphi \int_0^{\pi/2} d\vartheta \sin^k\vartheta F^2\left(\frac{\omega_z \cos\vartheta}{s_\alpha}\right) e^{\omega_z^2 l^2 \cos^2\vartheta / 2s_\alpha^2} \left\{(eA_{q\alpha})^2 + \frac{\omega_z^2}{s_\alpha^2} \delta_{j,L} \Xi_0^2\right\} \quad (2)$$

The parameter definitions are provided in the Appendix. The cubic Rashba coefficient depends on the HH-LH energy separation (Δ) and can be tuned by adjusting the strain in the SiGe layer, which is achieved by varying the composition x of the $Si_{1-x}Ge_x$ buffer layer. For SiGe/Ge, the large separation (Δ~110 meV) results in a smaller Rashba coefficient, influencing the decoherence dynamics.

The spin relaxation rate (1/$T_1$) as a function of strain for SiGe/Ge and GeSn/Ge QDs at B=0.5 and 1 T is shown in Figure 9. In SiGe/Ge, HH states dominate under compressive strain, resulting in a relaxation time of approximately 100 μs. As strain increases, LH states become more prominent due to the strain-induced band structure changes, leading to shorter $T_1$. This behavior is primarily due to the lower g-factor of LH compared to HH states, which decreases the Zeeman splitting and alters the interaction with phonons.

Within the LH-dominated regime, as strain further increases, the g-factor grows, raising the Zeeman energy. This enhanced Zeeman splitting increases the overlap between spin state transitions and the phonon density of states, facilitating phonon-mediated relaxation and reducing $T_1$. The elevated relaxation rate for LH compared to HH is attributed to the stronger SOC parameters associated with LH states. Additionally, the cubic Rashba term amplifies the sensitivity of LH relaxation rates to both strain and magnetic field variations. The peak in $1/T_1$ observed in Figure 9 corresponds to a strong anti-crossing in the energy levels caused by mixing between LH and HH states, where their interaction temporarily enhances the relaxation.

In the GeSn structure, LH states occupy the ground state while HH states are located at the barrier edges, resulting in low relaxation rates for LH in this configuration. At higher strain, the g-factor increases, further enhancing the overlap between spin state transitions and the phonon density of states, leading to an increased relaxation rate. The spin relaxation rate shows a $B^3$ dependence at low fields, transitioning to a $B^7$ trend at fields above approximately 0.1 T, as described in studies of SOC-mediated spin relaxation in quantum dots.[38-40]

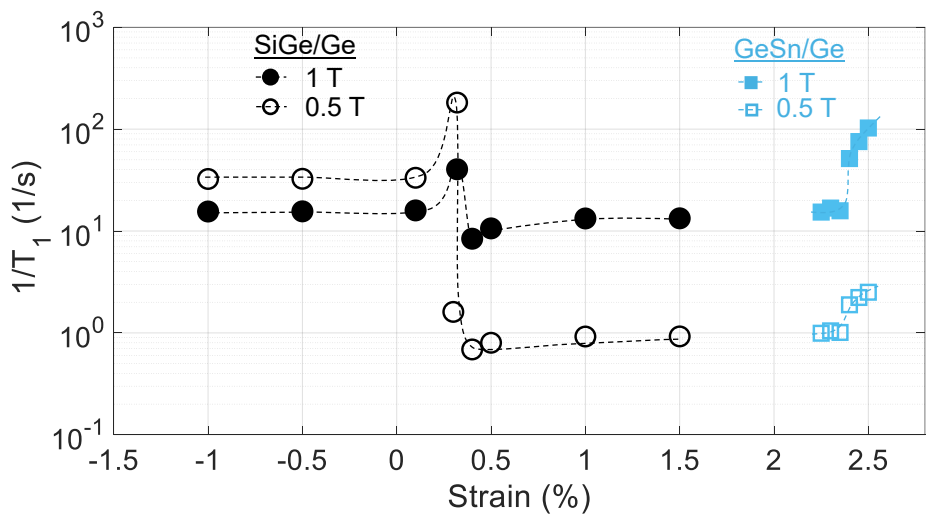

Figure 9: Spin relaxation rate ($1/T_1$) as a function of strain for SiGe/Ge and GeSn/Ge quantum dots at two magnetic field values (B=0.5 T and B=1 T). The sharp increase in $T_1$ observed for SiGe/Ge is attributed to the strong anti-crossing between LH and HH states near the critical strain, where HH-LH mixing is maximized. The relaxation rate is influenced by the SOC of phonon modes to the wavefunctions, with contributions from cubic Rashba SOC. For SiGe/Ge, the spin relaxation time remains relatively constant for HH and LH states across the strain regime, except near the anti-crossing. In the GeSn/Ge structure, where LH states dominate, the relaxation time increases steadily with strain due to reduced HH-LH mixing and weaker SOC-mediated phonon coupling.

### III.VI. DEPHASING

Charge noise in quantum dot qubits arises from the influence of electric field fluctuations on the g-factor. These fluctuations, often caused by electronics, defects, or nearby heterointerfaces, alter the qubit energy through spin-orbit coupling[41-43]. This strong sensitivity to electric fields is a defining characteristic of hole spin qubits: while a high g-factor enhances electrical tunability, it also increases susceptibility to electrical noise. Even small field variations can significantly impact the coherence and stability of hole spin qubits, leading to notable shifts in qubit behavior.

Charge noise generally exhibits a $1/f$ dependence, with higher noise power at lower frequencies. The magnitude of charge noise also varies significantly between materials and across different samples, depending on factors such as fabrication quality and material interfaces. Studies using methods like Single Electron Transistor (SET) spectroscopy have characterized these noise levels extensively[44-49].

The g-tensor for hole qubits shows substantial anisotropy due to asymmetries in both the confinement potential and the heterostructure. Using the model described in [50], we calculated the g-factor tensor by varying the in-plane direction of the magnetic field $B = (B\cos\phi, B\sin\phi, 0)$, where $\phi = [0, \pi/4, \pi/2]$. The variation of the g-factor with magnetic field direction is provided in the Appendix. The charge noise follows a $1/f$ spectrum with a noise spectral density given by $S(f) = A/|f|$, where A is the noise magnitude [41].

The dephasing rate, derived from charge noise, is calculated using the equation[51] [42]:

$$\frac{1}{T_2^*} = |\partial_{E_z}\omega|\sqrt{A\log(1/2\pi f_{ir}t)} \tag{3}$$

where 450 V²/m², $f_{ir}$=1 Hz, t=10 μs, and the qubit energy is given by $\hbar\omega = \sqrt{\mu_B^2 B_j B_k g_{ji} g_{ki}}$.

Figure 10 presents the calculated dephasing rates for the SiGe and GeSn systems. In SiGe, at compressive strain, HH states exhibit a low dephasing rate due to their lower g-factor. As strain increases, a peak is observed at the mixing point of LH and HH states, corresponding to enhanced susceptibility to charge noise. In the LH-dominated regime, the dephasing time improves by an order of magnitude compared to HH. The symmetric wavefunction in the xy-plane results in a symmetric g-factor tensor, causing the dephasing times for different magnetic field directions to overlap.

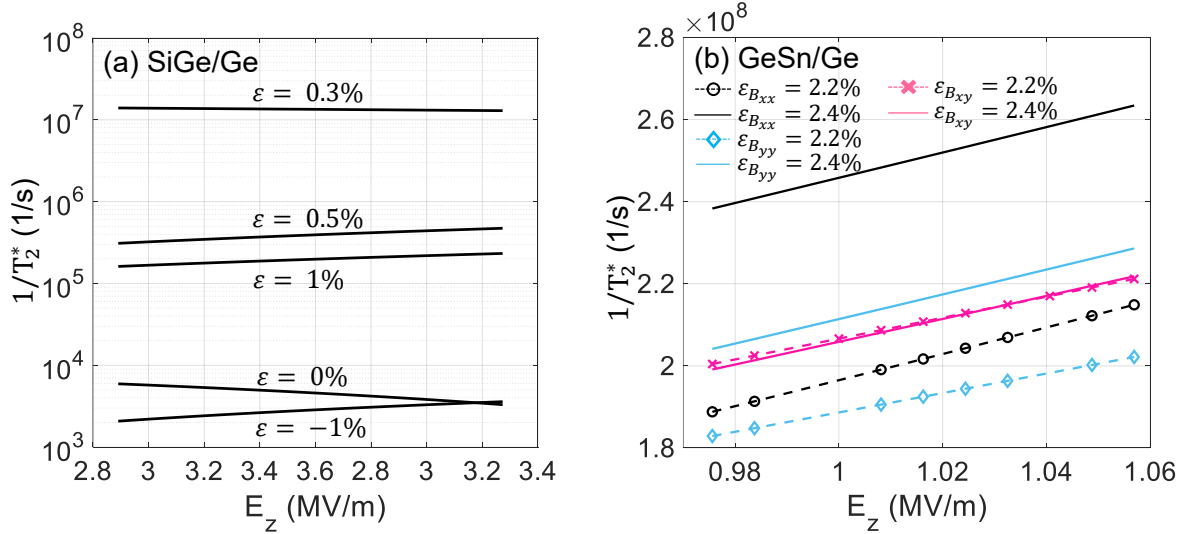

Figure 10: Dephasing rate ($1/T_2^*$) calculated for (a) SiGe and (b) GeSn systems as a function of the bias voltage $E_z$. In the SiGe system, the dephasing rate remains the same in all magnetic field directions due to the perfectly symmetric wavefunction within the quantum dot. The dephasing rate is lower for HH compared to LH, attributed to the transverse g-factor for HH states. For the xy direction, the dephasing rate decreases with increasing $E_z$, while in the x and y directions, it shows an increase with $E_z$.

In the GeSn system, the dependence of dephasing times on the magnetic field direction at different strain levels arises from the asymmetric wavefunction within the quantum dot. This asymmetry gives rise to an anisotropic g-factor in the xy-plane, as detailed in the Appendix, which causes variations in dephasing times depending on the direction of the applied magnetic field. Such anisotropy has also been observed in silicon-based qubits, where predictions of dephasing times were made by modeling "sweet spots" using higher-order terms in the Luttinger-Kohn Hamiltonian.[50] Experimentally, these sweet spots have been demonstrated in Si FinFETs, with dephasing times varying by a factor of two depending on the magnetic field direction.[52] Figure 10 illustrates the dephasing times for strain levels of 2.2% and 2.4%, highlighting the distinct variations resulting from this anisotropy in GeSn quantum dots.

## IV. Conclusion

This study investigates and compares two existing heterostructure systems, SiGe/Ge and GeSn/Ge, for their application in hole spin qubits, focusing on how strain influences critical parameters such as energy states, g-factor, Rabi frequency, spin relaxation, and dephasing times. The use of strain as a tunable parameter provides a powerful tool to differentiate the behavior of holes and analyze its effects on qubit performance metrics, offering insights into the design of robust quantum dot-based qubits.

Our results demonstrate that strain has a profound impact on the energy states in these systems, particularly in modulating the transition between HH and LH dominance. In SiGe/Ge quantum dots, compressive strain stabilizes HH states, while increasing strain induces a transition to LH-dominated states. This transition is marked by a peak in relaxation and dephasing rates due to strong HH-LH mixing. Beyond the critical strain, LH states dominate, exhibiting lower relaxation rates and improved coherence properties compared to HH states, attributed to their smaller g-factor and symmetric wavefunction in the quantum dot.

In GeSn/Ge systems, LH states dominate throughout the strain range due to the intrinsic tensile strain of the heterostructure. The relaxation rates calculated at two magnetic field strengths, 0.5 T and 1 T, show a higher relaxation rate at the higher field, consistent with the increased Zeeman splitting and enhanced phonon coupling. The dephasing rate in GeSn/Ge exhibits directional variability due to the anisotropic wavefunction and g-factor, reflecting the sensitivity of these systems to electric field fluctuations.

The study also highlights the relationship between strain and the Rabi frequency in these systems. In SiGe/Ge, the Rabi frequency peaks at the critical strain due to HH-LH mixing and stabilizes in the LH-dominated regime, while in GeSn/Ge, the Rabi frequency remains consistent across strain conditions but lower than that in SiGe/Ge. This indicates that LH qubits in Ge quantum dots combine the advantages of lower relaxation rates and higher Rabi frequencies, addressing current challenges in gate-defined spin qubits. By leveraging strain to tune critical parameters, our findings offer practical design strategies to enhance qubit coherence, manipulation, and scalability in quantum dot-based quantum technologies.

A significant difference observed between HH and LH spin qubits is in the anisotropy of their g-factors. For HH qubits, the out-of-plane g-factor ($g_\perp$) is larger than the in-plane g-factor ($g_\parallel$), while for LH qubits, the reverse is true, as observed for both LHs in GeSn/Ge and SiGe/Ge quantum dots. This difference directly influences the coupling mechanisms and sensitivity to electric and magnetic fields, further shaping their operational characteristics.

The choice between HH and LH spin qubits depends on the application requirements. HH spin qubits excel in coherence and robustness, making them ideal for systems prioritizing stability over manipulation speed. LH spin qubits, with their enhanced tunability and fast electric-field-driven operations, are better suited for applications where rapid control and scalability are paramount, provided decoherence can be mitigated through careful material and device design.

## V. Acknowledgment

This study was supported by Air Force Office of Scientific Research (AFOSR) and the Laboratory for Physical Sciences (LPS) under contract numbers FA9550-23-1-0302 and FA9550-23-1-0763. DV expresses gratitude to Nanoacademic Technologies Inc. for providing extended trial access to the QTCAD software.

## VI. APPENDIX

### A: k.p Parameters Used in the QD Simulations

| Parameters | Ge | Ge$_{0.89}$Sn$_{0.11}$ | Si$_{0.20}$Ge$_{0.80}$ | Refs |
|---|---|---|---|---|
| Luttinger-Kohn Parameters [$\gamma_1, \gamma_2, \gamma_3, \Delta$] | [13.38, 4.24, 5.69, 0.290eV] | [17.05, 6.249, 7.686, 0.352eV] | [6.1040, 1.1192, 2.2948, 0.0933] | 53,14,54 |
| Electron Effective Mass | 0.216m$_e$ | 0.0657 m$_e$ | 0.23m$_e$ | |
| Hole Effective mass | $m_{HH}$ = 0.34624157 $m_{LH}$ = 0.04230065 | $m_{HH}$ = 0.34624157 $m_{LH}$ = 0.04230065 | $m_{HH}$ = 0.34624157 $m_{LH}$ = 0.04230065 | 12,51 |
| Hole Degeneracy | 4 | 4 | 4 | |
| Valence band deformation potential | $a_v$ = 2.0 eV $b_v$ = −2.16 eV, $d_v$ = −6.06 eV | $a_v$ = 1.28 eV $b_v$ = −2.84 eV, $d_v$ = −5.15 eV | $a_v$ = 2.0 eV $b_v$ = −2.16 eV, $d_v$ = −6.06 eV | 50 |
| Hole g-factor [$\kappa, q$] | [3.41,0.06] | [1.7325, 0.0864] | [0.346,0.02] | 12,52 |
| Electron affinity ($\chi$,eV) | 4.05 | 3.5 | 4.05 | 52 |
| $E_g$ (0K, eV) | 0.67 | 0.31 | 0.858 | 50 |
| Relative Permittivity ($\varepsilon$) | 16 | 16.88 | 15.55 | |

### B. Energy Levels in SiGe/Ge QDs

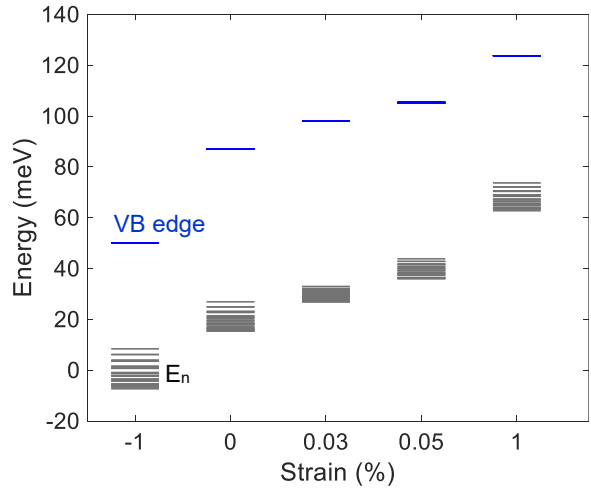

Figure A1: Energy levels for the quantum dot in the SiGe/Ge structure as a function of strain.

## C. Energy levels in GeSn/Ge QDs

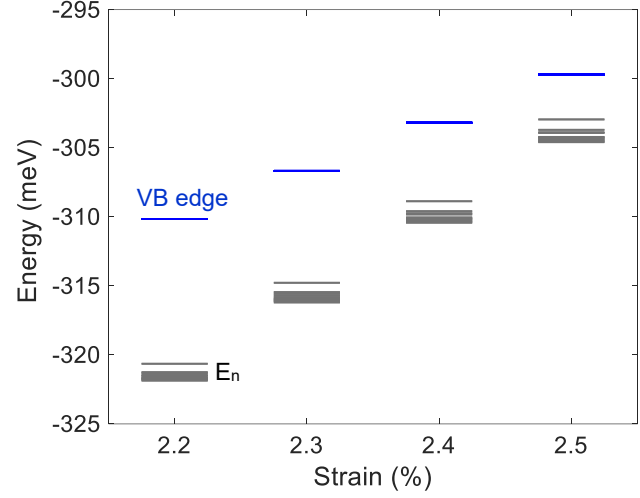

Figure A2: Energy levels for the quantum dot in the GeSn/Ge structure as a function of strain.

## D: Parameters Used in Decoherence Calculations

| Parameters | Definition | Formula | SiGe | GeSn | Ref |
|---|---|---|---|---|---|
| $\alpha$ | | $\alpha_R \langle E_z \rangle$ | $-2.377 \times 10^{54}$ | $-2.333 \times 10^{54}$ | |
| $\alpha_R$ | Rashba Coefficient | $\dfrac{-3e\hbar^3 \gamma_3^2}{2m_0^2 (E_{HH} - E_{LH})^2}$ | | | 32 |
| $\langle E_z \rangle$ | Electric field | - | $5.3 \times 10^6$ V/m | $13.3 \times 10^6$ V/m | |
| $m_0$ | Electron rest mass | - | $9.1 \times 10^{-31}$ | $9.1 \times 10^{-31}$ | |
| $\Delta$ | QD sub-band energy splitting | - | 5.4 meV | 0.7 meV | |
| $\rho$ | Mass density | - | 5323 kg/m$^3$ | 5323 kg/m$^3$ | |
| $\omega_z$ | Zeeman frequency | $(E_2 - E_1)/\hbar$ | $1.8477 \times 10^{12}$ rad/s | $2 \times 10^{12}$ rad/s | |
| $\omega_0$ | Characteristic Confinement Frequency | $(E_3 - E_1)/\hbar$ (without Field) | $1 \times 10^{13}$ | $1 \times 10^{12}$ | |
| $\omega_c$ | Cyclotron frequency | $|e|B/m^*c$ | $5.8550 \times 10^{11}$ | $4.18 \times 10^{12}$ | |
| $\omega_+$ | Harmonic oscillator frequency | $\Omega + \omega_c/2$ | $1.030 \times 10^{13}$ | $4.442 \times 10^{12}$ | |
| $\omega_-$ | Harmonic oscillator frequency | $\Omega - \omega_c/2$ | $9.7154 \times 10^{12}$ | $2.600 \times 10^{11}$ | |
| $\Omega$ | Frequency used in $\omega_\pm$ | $\sqrt{\omega_0^2 + \omega_c^2/4}$ | $1 \times 10^{13}$ rad/s | $2.35 \times 10^{12}$ rad/s | |
| $s_\alpha$ | Transverse and longitudinal phonon group velocities | - | [3350,4960] m/s | [3350,4960] m/s | |
| $l$ | Lateral size of QD | - | 50 nm | 90 nm | |
| $A_{q\alpha}$ | Piezoelectric constant | $\xi_i \xi_j$ | 0 | 0 | |
| $\Xi$ | Deformation potential | - | 15.2 eV | 15.2 eV | 55 |
| $F^2\left(\dfrac{\omega_z \cos\vartheta}{s_\alpha}\right)$ | Spread of the wavefunction | $F(k_z) = \int dz e^{ik_z z} |\psi_0(z)|^2$ | 1 | 1 | 37 |

## C: $g_{ij}$-Factors for Dephasing Rate Analysis

The calculation of the dephasing rate, as presented in the main text, depends on the variation of the g-factor tensor ($g_{ij}$). Figures A3 and A4 illustrate the $g_{xx}$, $g_{yy}$, and $g_{xy}$ components for the two heterostructures discussed.

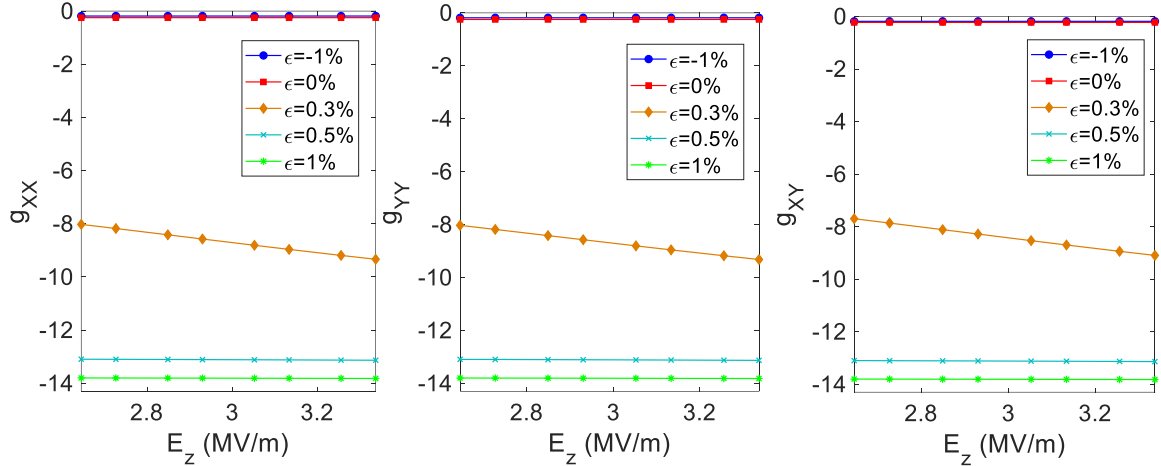

Figure A3: The g-tensor component for the SiGe/Ge structure. The g-factor is the same for every direction indicating a symmetric wavefunction in the x-y plane.

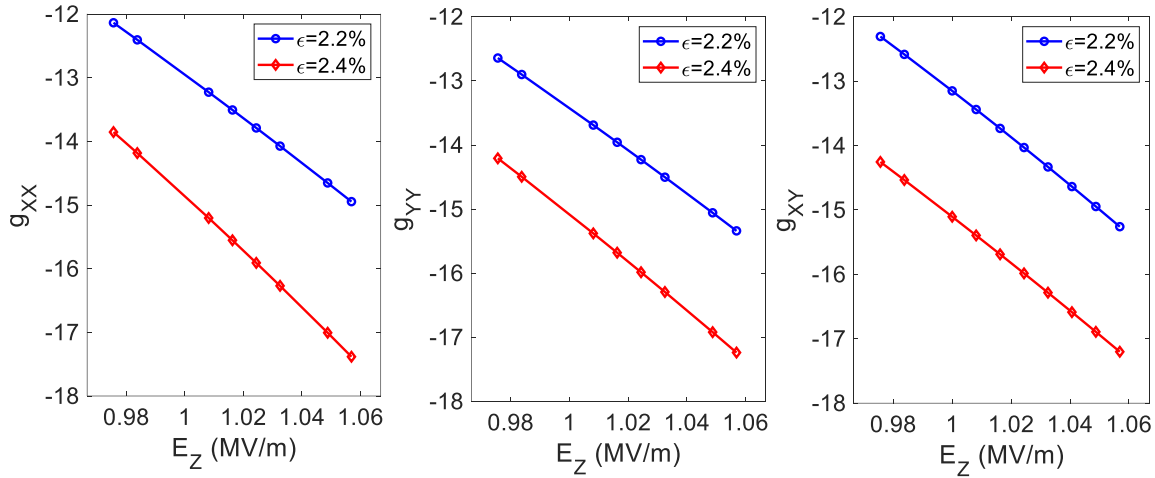

Figure A4: The g-tensor components for the GeSn/Ge structure. The slight variations in the g-factor are attributed to the asymmetry of the wavefunction within the xy-plain of the quantum dot.